\documentclass{nature}
\usepackage{graphicx}

\title{Differing Enceladean ocean circulation and ice shell geometries driven by tidal heating in the ice versus the core
}

\author{Wanying Kang$^{1\ast}$, Suyash Bire$^{1}$, Jean-Michel Campin$^{1}$, Christophe Sotin$^{2}$, Christopher German$^{3}$, 
Andreas Thurnherr$^{4}$ and John Marshall$^{1}$}

\begin{document}

\maketitle

\begin{affiliations}
 \item Earth, Atmospheric and Planetary Science Department, Massachusetts Institute of Technology, Cambridge, MA 02139, USA
 \item Jet Propulsion Laboratory, Caltech, 4800 Oak Grove Drive, Pasadena, CA, 91109, USA
 \item Woods Hole Oceanographic Institution, Woods Hole, MA 02543, USA
 \item Division of Ocean and Climate Physics, Lamont-Doherty Earth Observatory, Palisades, New York 10964, USA
\end{affiliations}

\begin{abstract}
  Beneath the icy shell encasing Enceladus, a small icy moon of Saturn, a global ocean of liquid water ejects geyser-like plumes into space through fissures in the ice, making it an attractive place to investigate habitability and to search for extraterrestrial life. The existence of an ocean on Enceladus has been attributed to the heat generated in dissipative processes associated with deformation by tidal forcing. However, it remains unclear whether that heat is mostly generated in its ice shell or silicate core. Answering this question is crucial if we are to unravel patterns of ocean circulation and tracer transport that will impact both the habitability of Enceladus and our ability to interpret putative evidence of any habitability and/or life. Using a nonhydrostatic ocean circulation model, we describe and contrast the differing circulation patterns and implied ice shell geometries to be expected as a result of heating in the ice shell above and heating in the core below Enceladus' ocean layer. If heat is generated primarily in the silicate core we would predict enhanced melting rates at the equator. In contrast, if heat is primarily generated in the ice shell we would infer a poleward-thinning ice geometry consistent with Cassini Mission observations.
\end{abstract}

Despite its small size (only 252 km in radius) and hence rapid heat loss, Enceladus retains a 40 km deep global ocean beneath its ice shell \cite{Thomas-Tajeddine-Tiscareno-et-al-2016:enceladus}. Geyser-like jets are ejected into space over the south polar region \cite{Porco-Helfenstein-Thomas-et-al-2006:cassini, Howett-Spencer-Pearl-et-al-2011:high, Spencer-Howett-Verbiscer-et-al-2013:enceladus, Iess-Stevenson-Parisi-et-al-2014:gravity}, providing a unique opportunity to peek through the 20km-thick ice shell\cite{Beuthe-Rivoldini-Trinh-2016:enceladuss, Hemingway-Mittal-2019:enceladuss} into the ocean below. Several flybys accomplished by the Cassini mission have produced vast amounts of data, greatly improving our understanding of this small but active moon. Particles and gases sampled from these jets, indicate the presence of organic matter \cite{Postberg-Khawaja-Abel-et-al-2018:macromolecular}, silica nanoparticles \cite{Hsu-Postberg-Sekine-et-al-2015:ongoing} and a modestly high pH environment \cite{Glein-Baross-Waite-2015:ph}, all suggestive of seafloor hydrothermal activity and astrobiological potential in the overlying ocean \cite{Glein-Postberg-Vance-2018:geochemistry, Taubner-Pappenreiter-Zwicker-et-al-2018:biological, McKay-Davila-Glein-et-al-2018:enceladus}. Gravity and topography measurements, combined with models of the interior, reveal Enceladus's cryosphere to be a poleward-thinning ice shell filled with ridges, scarps and large fissures, with cracks over the south polar region, where the ice shell is 15-20~km thinner than at the equator. \cite{Iess-Stevenson-Parisi-et-al-2014:gravity, Beuthe-Rivoldini-Trinh-2016:enceladuss, Tajeddine-Soderlund-Thomas-et-al-2017:true, Cadek-Soucek-Behounkova-et-al-2019:long, Hemingway-Mittal-2019:enceladuss}.

The spatially-variable ice shell thickness provides strong evidence for localized heat input given the tendency of ice flow \cite{Tobie-Choblet-Sotin-2003:tidally, Barr-Showman-2009:heat, Ashkenazy-Sayag-Tziperman-2018:dynamics}, the ice pump mechanism \cite{Lewis-Perkin-1986:ice} and more efficient heat loss through thinner ice shells \cite{Beuthe-2019:enceladuss, Kang-Flierl-2020:spontaneous}, to smooth out such heterogeneity. Heat is likely generated both in the ice shell itself \cite{Ojakangas-Stevenson-1989:thermal} and in the silicate core \cite{Choblet-Tobie-Sotin-et-al-2017:powering} as they are flexed by tidal forces. In contrast, no significant heating is expected from dissipation in the ocean  \cite{Chen-Nimmo-2011:obliquity, Beuthe-2016:crustal, Hay-Matsuyama-2019:nonlinear, Rekier-Trinh-Triana-et-al-2019:internal}. Which heat source dominates is still under debate, not least because of uncertainties associated with the assumed rheology and thus the rates and patterns of tidal heat generation of both the ice shell and the silicate core.
Hydrogen and nanometre-sized silica particles have been detected, providing clear geochemical evidence for active seafloor venting \cite{Hsu-Postberg-Sekine-et-al-2015:ongoing, Waite-Glein-Perryman-et-al-2017:cassini}. This submarine hydrothermalism is unlikely to be the dominant heat source, however, because it is insufficient to prevent gradual freezing of Enceladus' ocean \cite{Travis-Schubert-2015:keeping}. Tidal dissipation in the core, on the other hand, provides for a potentially strong and persistent source of heat at the bottom of the ocean \cite{Choblet-Tobie-Sotin-et-al-2017:powering}.
At the top of the ocean, reconstruction of ice shell geometry appears to be consistent with the heat being primarily generated in the ice shell \cite{Hemingway-Mittal-2019:enceladuss}. However, present dynamic models of the ice are unable to reproduce the heating rates required to maintain such a thin ice shell  \cite{Beuthe-2018:enceladuss, Beuthe-2019:enceladuss, Soucek-Behounkova-Cadek-et-al-2019:tidal}; attempts to boost heat generation with more advanced ice rheologies have been proposed \cite{McCarthy-Cooper-2016:tidal, Renaud-Henning-2018:increased, Gevorgyan-Boue-Ragazzo-et-al-2020:andrade}.

Whether heat is mostly generated in the ice or in the core has a leading order effect on ocean circulation and tracer (nutrition, biosignatures etc.) transport, which are expected, in turn, to affect both the habitability of Enceladus and our ability to detect such habitability and potential life. Moreover, as we shall see, the circulation itself has implications for the geometry of the ice shell, because it can concentrate heat and deliver it to the ice.  In the absence of direct measurements, here we explore how knowledge of ice shell geometry and ocean circulation patterns can provide new information on the partitioning of tidal heat generation between the ice shell and the core.

We consider two end-member scenarios: 1) all heat is assumed to be generated in the silicate core and transported upward by the ocean circulation (hereafter heat-from-below scenario) and 2) all heat is generated in the ice shell, directly compensating the heat loss through the ice shell due to heat conduction (hereafter heat-from-above scenario). Beginning from a 20~km-thick flat ice sheet \cite{Beuthe-Rivoldini-Trinh-2016:enceladuss}, we explore the ensuing ice shell melting patterns that develop if heating is imposed from above compared to below.

\begin{figure}[htbp!]
    \centering \includegraphics[width=0.8\textwidth]{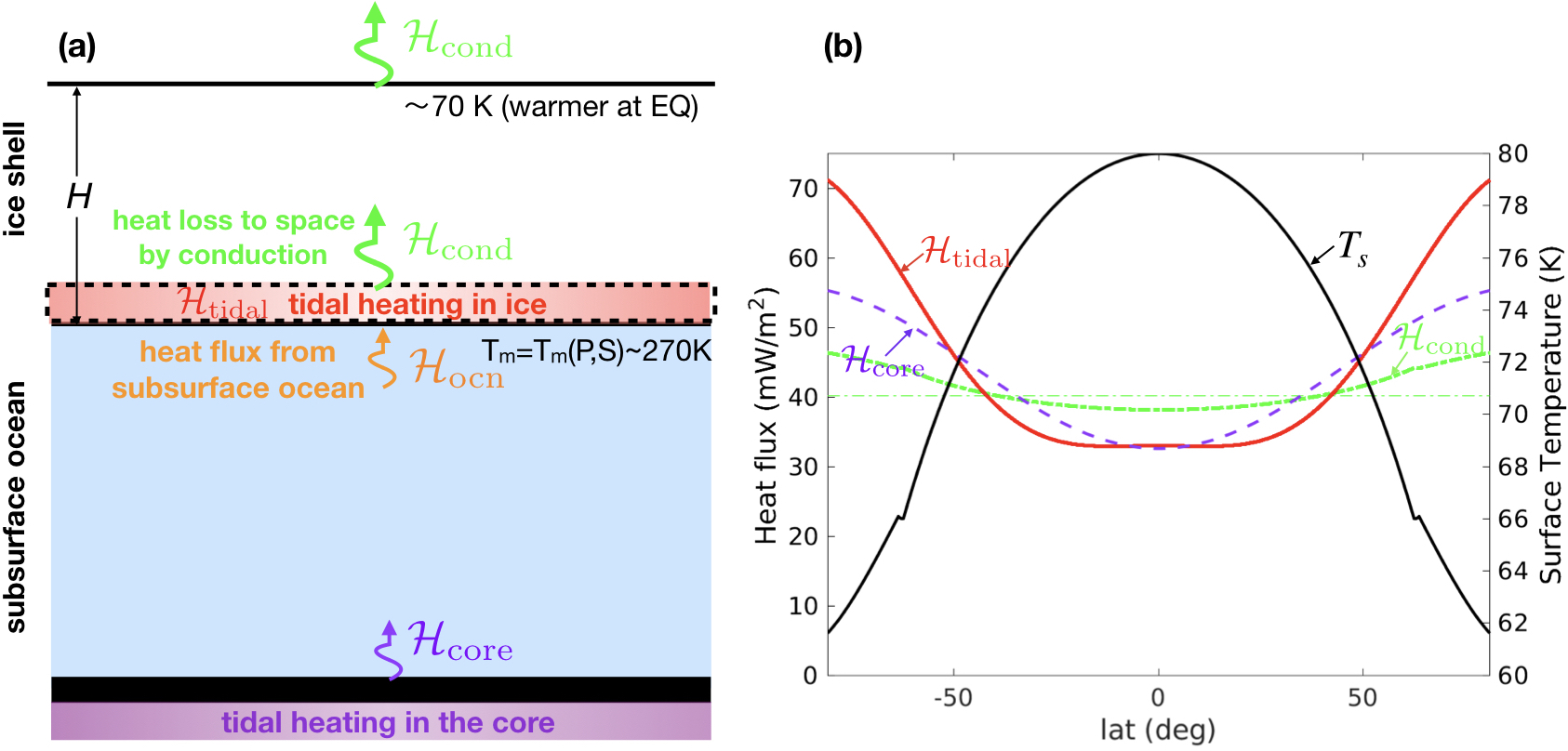}
    \caption{\small{The key physical processes incorporated into our model of Enceladus as discussed in detail in the Methods section. As sketched in panel (a), the ice shell freezing/melting rate is determined by the heat budget of a thin layer close to the liquid-ice interface (between the black dashed lines). All tidal heating of the ice is assumed to be generated here at the base of the ice shell, since the ice there is less elastic and more viscous \cite{Beuthe-2019:enceladuss}. The energy deficit of the layer equals the upward conductive heat flux $\mathcal{H}_{\mathrm{cond}}$ (green curly arrows) minus the heat carried upward by the ocean $\mathcal{H}_{\mathrm{ocn}}$ (orange curly arrow) and the generation of tidal heating within the box $\mathcal{H}_{\mathrm{tidal}}$ (red shading). Heating from the silicate core (purple shading) at the bottom of the ocean is denoted by the purple curly arrow. 
    The meridional profiles of heat fluxes and surface temperature are shown in panel (b). Both the dissipation rate in the ice ($\mathcal{H}_{\mathrm{tidal}}$, red solid line) and the core ($\mathcal{H}_{\mathrm{core}}$, purple dashed line) peak at the poles (but the gradient of  $\mathcal{H}_{\mathrm{tidal}}$ is greater), as does the heat loss through the ice shell due to conduction ($\mathcal{H}_{\mathrm{cond}}$, thick green dash-dotted line). The global mean values for $\mathcal{H}_{\mathrm{tidal}}$ and $\mathcal{H}_{\mathrm{core}}$ are rescaled so that their global mean equals that of the heat conduction rate $\overline{\mathcal{H}_{\mathrm{cond}}}$, which is plotted as a thin green dash-dotted line. We set the surface temperature $T_s$ (black solid line) using an empirical formula given by \textit{Ojakangas \& Stevenson 1989}\cite{Ojakangas-Stevenson-1989:thermal} based on radiative equilibrium. The equatorial region receives more solar radiation and is therefore warmer than at high latitudes.}}
    \label{fig:schematics}
  \end{figure}

The key processes included in our model are sketched in Fig.~\ref{fig:schematics}(a). The heat generated in the silicate core $\mathcal{H}_{\mathrm{core}}$ (purple curly arrow) is redistributed and transmitted to the water-ice interface by the ocean circulation, as denoted by an orange curly arrow. This $\mathcal{H}_{\mathrm{ocn}}$, together with the tidal heat generation in the ice shell $\mathcal{H}_{\mathrm{tidal}}$ (red shading), compensates heat conduction through the ice shell, $\mathcal{H}_{\mathrm{cond}}$, and thence radiation to space; the residual leads to melting or freezing at the base of the ice shell. We simulate the ocean dynamics using MITgcm \cite{MITgcm-group-2010:mitgcm, Marshall-Adcroft-Hill-et-al-1997:finite} configured for Enceladus in which non-hydrostatic effects and full treatment of Coriolis forces are included and the shallow fluid approximation relaxed. In a departure from previous studies \cite{Soderlund-Schmidt-Wicht-et-al-2014:ocean}, we impose a heat flux at the bottom rather than specify a temperature contrast. 
Further, the interaction between ice and ocean is represented by a modified version of MITgcm's ``shelf-ice'' package, in which both the meridional variations of tidal heat generation in the ice and in the core, as well as the ice surface temperature are accounted for (the corresponding profiles are plotted in Fig.~\ref{fig:schematics}b; see methods for technical details). As discussed in the methods section, our simulations are in the rapidly rotating convection regime as reviewed by Gastine et al.\cite{Gastine-Wicht-Aubert-2016:scaling}


\begin{figure}[htbp!]
    \centering
\includegraphics[width=\textwidth]{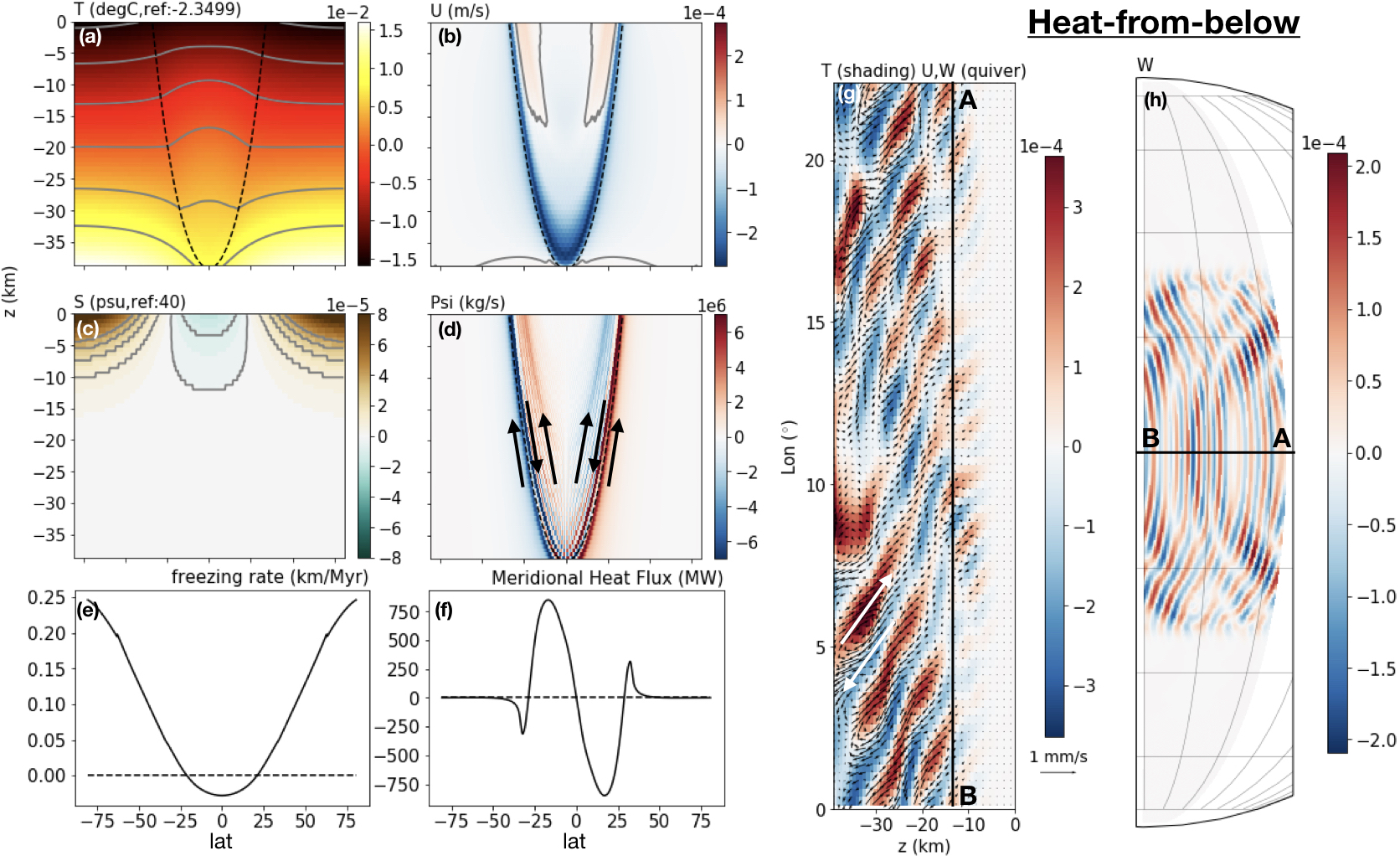}
    \caption{\small{Thermodynamic and dynamic fields from the ``heat-from-below'' scenario. Panels (a-d) show the zonally-averaged temperature $T$, zonal flow speed $U$, salinity $S$ and meridional streamfunction $\Psi$ (blue color denotes clock-wise circulation). The orientation of the tangent cylinder is marked by dashed curves. The vertical axis spans the whole $39.2$~km ocean, and the origin is set at the interface between ice and ocean. The zero speed contours are plotted in panel (b) using gray solid lines to mark the transition between easterlies to westerlies. Isothermal and isosaline contours are plotted in panels (a,c) to highlight the meridional gradients. Panels (e,f) shows the zonally-averaged freezing rate and the vertically and zonally integrated meridional heat transport as a function of latitude. The dynamics of the ``equatorial rolls'' is presented in panels (g,h). Panel (g) shows the equatorial vertical section of temperature $T$ (shading), zonal speed $U$ and vertical speed $W$ (arrows). Panel (h) shows a horizontal plot of $W$ at the level marked by a solid black line in panel (g).} }
    \label{fig:heat-from-below}
  \end{figure}

  When heated from below, the heat generated in the silicate core is carried away into the interior of the ocean. Although the prescribed heat generation rate is slightly lower at the equator than at the poles \cite{Beuthe-2019:enceladuss}, the upward heat transport in the tropics is much more efficient than at higher latitudes (see below). The signature of more efficient vertical heat transport is the weaker vertical temperature gradient near the equator (see Fig.~\ref{fig:heat-from-below}a), consistent with findings by \textit{Aubert 2005}\cite{Aubert-2005:steady} in the context of the dynamo of Earth's iron core. As a result, the upper ocean becomes warmer at low latitudes, giving rise to melting of the ice shell there and freezing elsewhere (Fig.~\ref{fig:heat-from-below}e). (See \cite{Amit-Choblet-Tobie-et-al-2020:cooling} for a discussion of polar vs.\ equatorial concentration of bottom heating in numerical experiments applied to Titan). This would eventually reshape the ice shell into one that is thinner at the equator and thicker at the poles, in contrast to what has been reported for Enceladus \cite{Iess-Stevenson-Parisi-et-al-2014:gravity, Beuthe-Rivoldini-Trinh-2016:enceladuss, Tajeddine-Soderlund-Thomas-et-al-2017:true, Cadek-Soucek-Behounkova-et-al-2019:long, Hemingway-Mittal-2019:enceladuss}. The freshwater input (brine rejection) associated with the melting (freezing) causes a reduction (increase) in the local salinity, as shown in Fig.~\ref{fig:heat-from-below}(c).

  The more efficient upward heat transport near the equator cools the abyss making the equator colder than higher latitudes in the lower part of the ocean. This meridional temperature gradient induces equatorward (downgradient) heat transport in the lower part of the ocean (Fig.~\ref{fig:heat-from-below}f), sustaining strong upward heat transport at low latitudes.

  The equatorial region is more efficient in transporting heat upward because of the formation of ``equatorial rolls'': convection constrained by the rotation vector which is largely perpendicular to local gravity. In Fig.~\ref{fig:heat-from-below}(g,h), the temperature and velocity anomalies associated with the rolls are shown in both zonal section  and plan view. The rolls emanate from the warm sea floor and tilt eastward (in the prograde direction) with height. Warm water rises and cold water sinks along these tilted trajectories -- as a result, heat is transported upward. These rolls fill the whole region in the low latitudes outside the tangent cylinder\footnote{The cylinder aligned with the moon's axis of rotation, and whose surface is tangential to the silicate core.}. This is the region in which the water column is the most efficiently warmed from below (Fig.~\ref{fig:heat-from-below}a). These characteristics are consistent with \textit{Cardin \& Olson 1994}\cite{Cardin-Olson-1994:chaotic} and \textit{Christensen 2002}\cite{Christensen-2002:zonal}, where the dynamics of such rolls have been demonstrated in laboratory experiments, explained by a mode growth in a bottom-heated fluid encased by a spherical shell, and reproduced in numerical simulations.
  
  The circulation is strongly constrained by rotation and the geometry of the tangent cylinder, as expected when perturbing a rapidly-rotating\footnote{The dominance of rotation is suggested by a small Rossby number, diagnosed here to be $3.2\times10^{-6}$.} incompressible fluid according to the Taylor-Proudman theorem. The bowl-shape structure in the zonal flow field (Fig.~\ref{fig:heat-from-below}b) and the streamfunction for the meridional overturning circulation (Fig.~\ref{fig:heat-from-below}d) are strongly aligned with the tangent cylinder (dashed curve). Moreover, because rotational effects are so dominant, the zonal current and temperature distribution are in thermal wind balance (not shown), a consequence of geostrophic and hydrostatic balance. As can be seen in Fig.~\ref{fig:heat-from-below}(a), in the upper ocean the temperature near the tangential cylinder is lower than outside the tangential cylinder (enclosed by the bowl-shaped curve). The reverse is true in the lower part of the ocean. By thermal wind this implies a zonal current that becomes more westward (eastward) on moving upward following the Taylor column which originates from slightly higher (lower) latitudes than the tangential cylinder (see Fig.~\ref{fig:heat-from-below}b). The meridional overturning circulation is clockwise (counter clockwise) in the high-latitude (low-latitude) flank of the tangential cylinder in the southern hemisphere. This circulation brings cold water downward near the tangential cylinder, giving rise to the aforementioned temperature profile.
  

\begin{figure}[htbp!]
    \centering
\includegraphics[width=\textwidth]{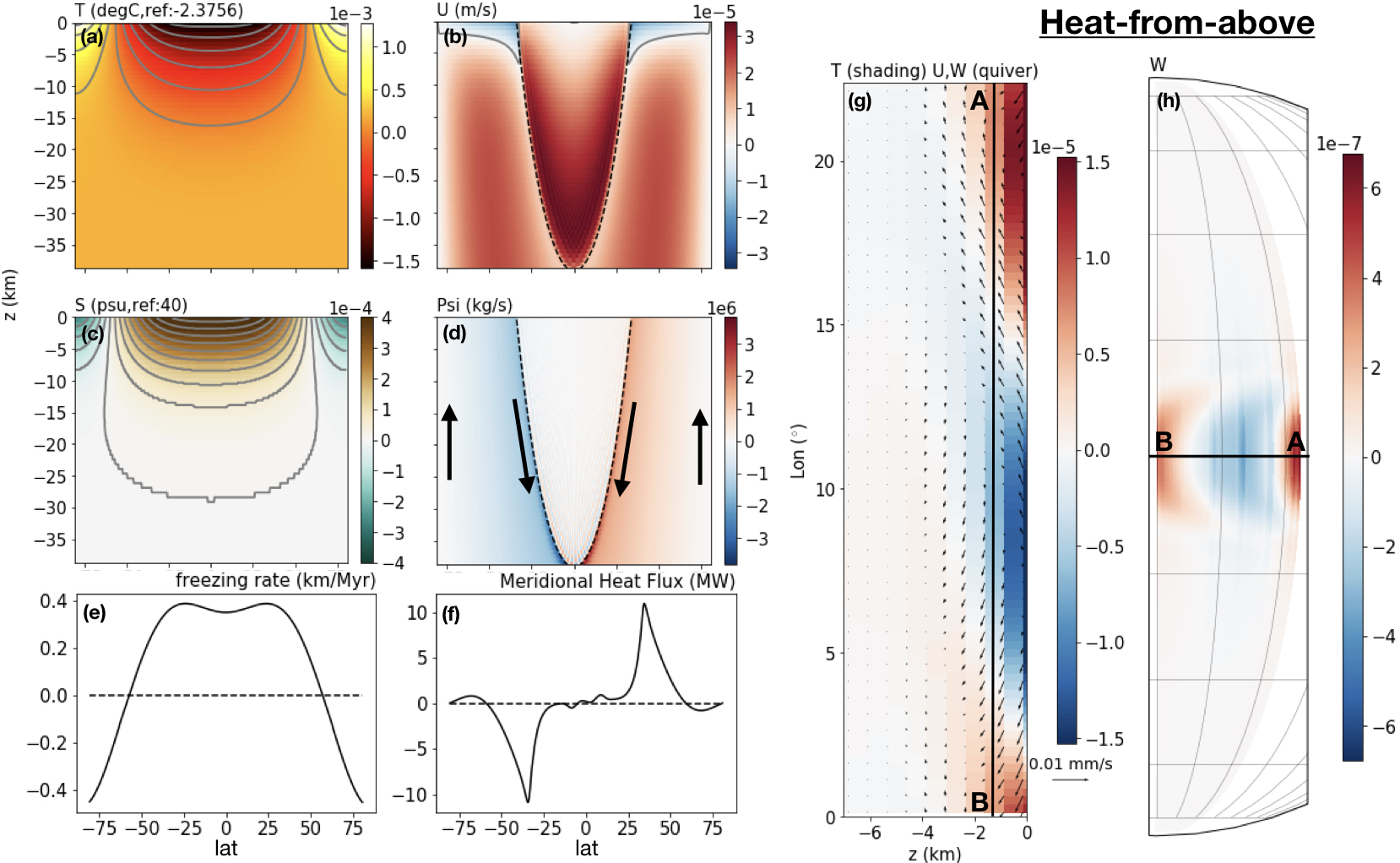}
    \caption{\small{Same as Fig.~\ref{fig:heat-from-below}, but for the ``heat-from-above'' scenario. To zoom in to the upper layer of the ocean, where the dynamics is active, we only show the top 7~km beneath the ice shell in panel (g), and we adjust the direction of arrows to reflect the changed aspect ratio of the domain. Also, the arrow sizes in panel (g) have been scaled up by a factor of 100 compared to Fig.~\ref{fig:heat-from-below} to better show the weak perturbations here.}}
    \label{fig:heat-from-above}
  \end{figure}

  The ocean circulation is very different in the ``heat-from-above'' scenario. With heat directly generated within the ice shell, wherever tidal heating dominates heat loss due to conduction the ice shell melts, and vice versa. Both heat generation in the ice and loss from the surface are polar amplified (see methods for details), but the meridional variation of the former is greater than that of the latter, causing melting at high latitudes and freezing at low latitudes (Fig.~\ref{fig:heat-from-above}e). Thus we  expect the ice shell on Enceladus to evolve toward a geometry that is thinner at the poles than at the equator, just as inferred from Cassini observations \cite{Iess-Stevenson-Parisi-et-al-2014:gravity, Beuthe-Rivoldini-Trinh-2016:enceladuss, Tajeddine-Soderlund-Thomas-et-al-2017:true, Cadek-Soucek-Behounkova-et-al-2019:long, Hemingway-Mittal-2019:enceladuss}. 

  Melting (freezing) near the poles (the equator) reduces (increases) the salinity near the ocean-ice interface (see Fig.~\ref{fig:heat-from-above}c), which in turn increases (decreases) the freezing temperature (Fig.~\ref{fig:heat-from-above}a). If this is indeed what is happening on Enceladus, we would expect its ocean to be saltier than the spray samples. The associated meridional density gradient drives a meridional overturning circulation in which there is sinking along the tangential cylinder (marked by a dashed curve) and water returning back to the surface at higher latitudes (Fig.~\ref{fig:heat-from-above}d). By invoking angular momentum conservation, equatorward motion in the upper ocean induces westward flow and poleward motion below induces eastward flow (Fig.~\ref{fig:heat-from-above}b), as seen in our solution. At high latitudes, equatorward motion carries warm water from the poles, transporting heat equatorward (Fig.~\ref{fig:heat-from-above}f): at middle latitudes, sinking of cold water along the tangent cylinder transports heat poleward (Fig.~\ref{fig:heat-from-above}f).

  This general pattern of circulation remains in place and strengthens in time during the $120$~years of our simulation, along with salinity anomalies that grow beneath the ice shell. By the end of the $120$-year simulation, only $0.1$~m of ice topography has been developed between the equator and the pole, and the corresponding mean oceanic upwelling at the poles is merely $10^{-8}$~m/s. However, over time, the salinity changes due to freezing and melting will accumulate as the ice shell gets reshaped over a timescale of millions of years, and we can expect salinity-driven meridional circulation to strengthen with it.

  Driven from above, ocean currents are an order of magnitude weaker than that in the ``heat-from-below'' scenario, consistent with Sandstrom's Theorem \cite{Sandstroem-1908:dynamische}. The predicted $T$ and $S$ anomalies are confined to the uppermost $5$~km (Fig.~\ref{fig:heat-from-above}g,h), a small fraction of the $40$~km deep ocean \cite{Hemingway-Iess-Tadjeddine-et-al-2018:interior}. Unlike when warming from below, the perturbation fields tilt westward (retrograde) with height, the vertical component of the motion is one order of magnitude weaker than the horizontal component (arrows in Fig.~\ref{fig:heat-from-above}g are mostly aligned with the y-axis), and they are dominated by the gravest mode in the zonal direction permitted by the domain size. The amplitude of the equatorial wave structure also grows with time as the ocean becomes more and more unstably stratified. 

\begin{figure}[htbp!]
    \centering
\includegraphics[width=0.85\textwidth]{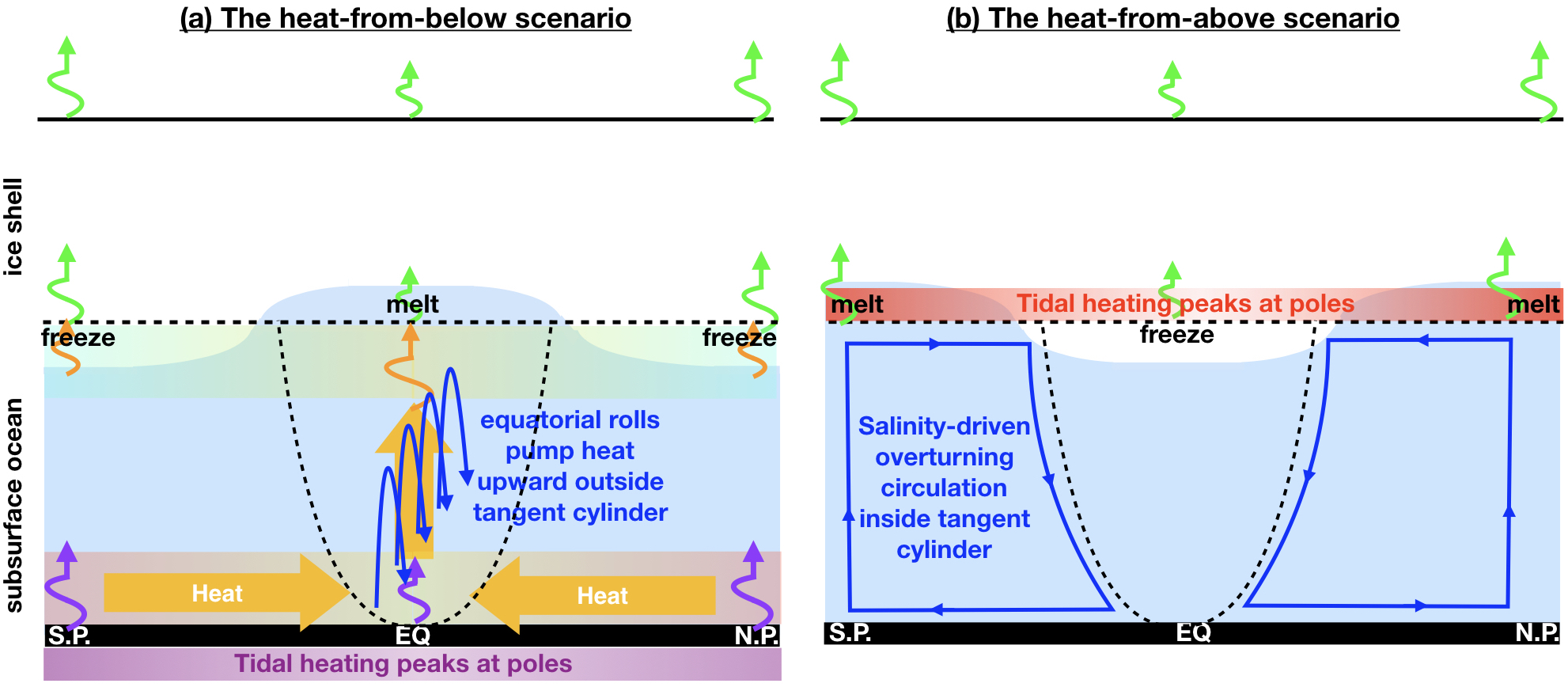}
    \caption{\small{Key aspects of our modelling. The notations of heat fluxes are the same as defined in Fig.~\ref{fig:schematics}a. Panel (a) demonstrates aspects of the ``heat-from-below'' solution. The ocean becomes destabilised by seafloor heating. Equatorial rolls (blue arrows) grow in the tropics out of the tangent cylinder (marked by a black dashed curve) and efficiently transport heat upward leading to ice melt at low latitudes. As heat is transported upwards (vertical orange arrow), the equatorial bottom water becomes cooler (shading in the bottom part of the ocean), inducing equatorward heat transport (horizontal orange arrow). Meanwhile, ice is formed at higher latitudes. 
      Panel (b) demonstrates key aspects of the ``heat-from-above'' scenario.
    A polar-amplified tidal heating profile leads to melting of ice at the poles. Salinity changes induced by polar melting and tropical freezing support a meridional overturning circulation (blue arrow). This scenario predicts a poleward-thinning of the ice shell consistent with observations.
     } }
    \label{fig:mechanism}
  \end{figure}

  To summarize, key outcomes from our study are sketched in Fig.~\ref{fig:mechanism}. The poleward-thinning of Enceladus's ice shell, as indicated by observations, \cite{Iess-Stevenson-Parisi-et-al-2014:gravity, Beuthe-Rivoldini-Trinh-2016:enceladuss, Tajeddine-Soderlund-Thomas-et-al-2017:true, Cadek-Soucek-Behounkova-et-al-2019:long} suggests a scenario in which heat is predominantly generated in the ice shell, consistent with inferences from ice geometry using an interior model \cite{Hemingway-Mittal-2019:enceladuss}. Forced by near-surface salinity variations due to long-term melting/freezing, upward motions gradually develop over the poles in the ``heat-from-above'' scenario. When heat is only supplied from the core, the ice shell tends to develop a poleward thickening geometry, even though the heating profile (slightly) peaks over the polar regions. In a scenario in which 50\% of the heat comes from the shell and 50\% comes from the core (not shown), we find that the shell still thins moving polewards. It should also be remembered that once significant ice thickness gradients are established, heat generation in the ice would be further amplified in places where the shell is thinner \cite{Beuthe-2019:enceladuss}. This effect, as argued by \textit{Kang \& Flierl 2020}\cite{Kang-Flierl-2020:spontaneous}, sustain the significant variations of ice thickness, as suggested by observations \cite{Iess-Stevenson-Parisi-et-al-2014:gravity, Tajeddine-Soderlund-Thomas-et-al-2017:true, Cadek-Soucek-Behounkova-et-al-2019:long, Hemingway-Mittal-2019:enceladuss}, in presence of ice flow and others mechanisms \cite{Tobie-Choblet-Sotin-2003:tidally, Barr-Showman-2009:heat, Ashkenazy-Sayag-Tziperman-2018:dynamics, Lewis-Perkin-1986:ice, Beuthe-2019:enceladuss} that tend to damp out the gradients.
  
  In conclusion we note that mean upwelling near the poles in our polar-thinning solution provides an advective pathway for dissolved and particulate materials (including those bearing potential biosignatures) to be transported from the seafloor to the ocean-ice interface, likely aided by localized convection driven by seafloor venting \cite{Speer-1997:thermocline}, and finally ejected into space as geysers.
  Although our results do not support a dominant role for the ``heat-from-below'' scenario, bottom heating that is concentrated through narrow regions, as suggested by \textit{Vance \& Goodman 2009}\cite{Vance-Goodman-2009:oceanography}, \textit{Choblet et al. 2017}\cite{Choblet-Tobie-Sotin-et-al-2017:powering} and \textit{Soucek et al. 2019}\cite{Soucek-Behounkova-Cadek-et-al-2019:tidal}, requires further study and could be addressed using the modeling framework set out here. The effects of ocean dynamics (especially when a significant amount of bottom heat is released over a small area), and how equilibrium could be reached through the coupling between the ocean and ice, are clearly key and need to be explored further.

\begin{methods}
  \section*{Model description.}
  The simulations are carried out using the state-of-the-art Massachusetts Institute of Technology OGCM (MITgcm \cite{MITgcm-group-2010:mitgcm, Marshall-Adcroft-Hill-et-al-1997:finite}). The model integrates the non-hydrostatic primitive equations for an incompressible fluid in height coordinates including a full treatment of the Coriolis force in a deep fluid. In particular it can capture the dynamics of convection in the presence of planetary rotation which has a projection on to the local horizontal. These components are typically ignored when simulating Earth's ocean because of its small aspect ratio (the ratio between depth and horizontal scale of the ocean basin), but is crucial for Enceladus's ocean, whose aspect ratio is instead order $40$km$/252$km$\sim0.16$ and so not small. Since the depth of the Enceladus's ocean is comparable to its radius, the variation of gravity with depth is significant. The vertical profile of gravity in the ocean and ice shell of Enceladus is given by, assuming a bulk density of $\rho_{\mathrm{out}}=1000$~kg/m$^3$:
  \begin{equation}
    \label{eq:g-z}
    g(z)=\frac{G\left[M-(4\pi/3)\rho_{\mathrm{out}}(a^3-(a-z)^3)\right]}{(a-z)^2}.
  \end{equation}
 In the above equation, $G=6.67\times10^{-11}$~N/m$^2$/kg$^2$ is the gravitational constant and $M=1.08\times 10^{20}$~kg and $a=252$~km are the mass and radius of Enceladus.

  Previous studies generally suggest a fresher ocean on Enceladus than that on Earth, however, due to the lack of direct measurements, there are great uncertainties. A $5$-$20$~psu (g/kg) salinity has been derived based on the composition of the E-ring particles, remnants of the south-pole geyser explosions \cite{Postberg-Kempf-Schmidt-et-al-2009:sodium}. To avoid freezing of the liquid–gas interface at the south pole, a salinity of at least $20$~psu is required, according to \cite{Ingersoll-Nakajima-2016:controlled}. Based on the existence of silica nano-particles in the plume fluid, Hsu et al. places an upper bound on the salinity of Enceladus' ocean at $40$~psu \cite{Hsu-Postberg-Sekine-et-al-2015:ongoing}. In this work, we initialize the salinity using this upper bound and adopt a linear equation of state (EOS, determines how density varies with temperature and salinity) to make the dynamics as transparent as possible. The thermal expansion coefficient and the haline contraction coefficient are set to the first derivative of density with respect to temperature and salinity using the Gibbs Seawater Toolbox \cite{McDougall-Barker-2011:getting}.

  The small size of Enceladus, together with our choice to model only a $22.5$~deg sector of the sphere, enables us to perform high resolution simulations. We use a one-quarter degree resolution corresponding to roughly 1~km resolution at the equator and 0.17~km resolution at the poles, comparable to that used in a recent study of Europa \cite{Ashkenazy-Tziperman-2020:europas}. Forced by a small heat flux (see Fig.1b), it takes 100s of years for the model to reach equilibrium. In the vertical direction, the $40$ km ocean is separated into $60$ layers and a $20$ km thick ice shell resides on top. The spacing between layers slightly decreases approaching the boundary layers at the top and bottom to better represent them. The size of each grid shrinks with depth due to the spherical geometry and is accounted for by switching on the ``deepAtmosphere'' option of MITgcm.

  Sub-grid scale mixing processes that are not resolved by the model are represented by explicit eddy diffusion and viscosity coefficients. We estimate the horizontal viscosity and diffusivity based on the scaling for geostrophically adjusted convection. According to \textit{Jones and Marshall 1993}\cite{Jones-Marshall-1993:convection}, convection plumes driven by bottom heating in a rapidly-rotating frame would coalesce to spinning ``cones'', whose size and speed scale as:
  \begin{eqnarray}
      l_{\mathrm{cone}}&\sim&\left(\frac{BD^2}{\Omega^3}\right)^{1/4}\\
      u_{\mathrm{cone}}&\sim&\left(BD^2\Omega\right)^{1/4},
  \end{eqnarray}
  where $D$ is the ocean depth and $\Omega$ is the rotation rate, $B$ is the buoyancy flux, which is related to the heat flux $Q$ by $B=\left.\alpha g Q\right/\rho C_p$, where $\alpha=4.27\times10^{-5}/K$ is the thermal expansion coefficient, $C_p=4000$~J/kg/K is the heat capacity of the ocean and $\rho=1000$~kg/m$^3$ is the ocean density. Adopting a $43$~mW/m$^2$ heat flux, we find $l_{\mathrm{cone}}=86.5$~m and $u_{\mathrm{cone}}=0.00925$~m/s. We set the horizontal viscosity $\nu_h$ and diffusivity $\kappa_h$ to be equal to the product of $l_{\mathrm{cone}}$ and $u_{\mathrm{cone}}$, which gives $\nu_h=\kappa_h=0.8$~m$^2$/s. Even lower viscosity has been tested ($\nu_h=\kappa_h=0.1$~m$^2$/s), and the thermodynamic field remains qualitatively unchanged but with some grid-point noise. We therefore adopted a $\nu_h=\kappa_h=1$~m$^2$/s.
  To maintain numerical stability, whilst avoiding dynamics being overridden by diffusion, we set the vertical diffusion and viscosity coefficient $\nu_r$, $\kappa_r$ to $0.01$~m$^2$/s, which is $2$ orders of magnitude smaller than that used in the horizontal direction.
  
  To compare to previous numerical studies \cite{Amit-Choblet-Tobie-et-al-2020:cooling, Gastine-Wicht-Aubert-2016:scaling}, we estimate the Rayleigh number comensurate with our experiments:
  \begin{equation}
      \mathrm{Ra}=\frac{\alpha g\Delta T D^3}{\nu_r \kappa_r}\sim 10^{11},
  \end{equation}
  where $\Delta T=40$~mK is the temperature difference between the diagnosed top and bottom temperatures, Enceledean gravity is $g=0.113$~m/s$^2$, rotation rate $\Omega=5.3\times 10^{-5}$s$^{-1}$ and ocean depth $H=40.2$~km. Our Ekman number $E$ estimate is:
  \begin{equation}
      E=\frac{\nu}{\Omega/D^2}\sim 10^{-7},
  \end{equation}
 where vertical diffusivity and viscosities have been used, $\nu_r=\kappa_r=0.01$~m$^2$/s. The resulting Rayleigh/Ekman number pairing places our ``heat-from-below'' scenario in the ``rapidly-rotating'' regime in Gastine's diagram \cite{Gastine-Wicht-Aubert-2016:scaling}, which is consistent with more heat going through the equatorial ice shell.

  \subsection{Heat balance.}
  
  On Enceladus, tidal heating generated in the silicate core $\mathcal{H}_{\mathrm{core}}$ and the ice shell $\mathcal{H}_{\mathrm{tidal}}$ should exactly compensate the heat loss through the ice shell due to conduction $\mathcal{H}_{\mathrm{cond}}$ in steady state.
  The expressions of $\mathcal{H}_{\mathrm{cond}}$ is adopted from the heat conduction model and the tidal heating model in \cite{Kang-Flierl-2020:spontaneous}, which is a simplified version from \cite{Beuthe-2019:enceladuss}. We assume the heat conductivity of ice to be inversely dependent on temperature, $\kappa=\kappa_0/T$, so that $\mathcal{H}_{\mathrm{cond}}$ is proportional to the difference of the logarithm of temperature at the top and bottom of the ice shell
  \begin{equation}
    \mathcal{H}_{\mathrm{cond}}=\frac{\kappa_{0}}{H} \ln \left(\frac{T_{\mathrm{ocn-top}}}{T_{s}}\right),\label{eq:H-cond}
  \end{equation}
  where $H=20.8$~km is the thickness of the ice shell and $T_s$ is temperature at the ice-air interface of the ice shell. We calculate $T_s$ using the approximation formula in \textit{Ojakangas and Stevenson 1989} \cite{Ojakangas-Stevenson-1989:thermal}, which is based on radiative equilibrium ignoring modulation induced by the tidal heating.
  \begin{eqnarray}
    T_s=&T_{s0}\cos^{1/4}\phi &,\phi<\pi/2-\phi_{\mathrm{obl}}\\
T_s=&T_{s0}\left[\left(\left.\phi_{\mathrm{obl}}^2+(\pi/2-\phi)^2\right)\right/2\right]^{1/8} &,\phi\geq\pi/2-\phi_{\mathrm{obl}}
  \end{eqnarray}
  where $T_{s0}=80$K is the temperature at the equator and $\phi_{\mathrm{obl}}=27$~deg is the obliquity of Saturn and Enceladus. The above formula yields a temperature profile that cools moving poleward with a global-mean surface temperature of roughly 70K (see the black curve in Fig.~\ref{fig:schematics}b).

  In the heat-from-above experiment, we set $\mathcal{H}_{\mathrm{tidal}}$ equal to the heat loss rate $\mathcal{H}_{\mathrm{cond}}$ and set $\mathcal{H}_{\mathrm{core}}$ to zero; while in the heat-from-below experiment, we allow $\mathcal{H}_{\mathrm{core}}$ to compensate $\mathcal{H}_{\mathrm{cond}}$. Note that the core has a smaller surface area than the water-ice interface, so instead of setting $\mathcal{H}_{\mathrm{core}}=\mathcal{H}_{\mathrm{core}}$, we need to multiply a factor of $\left.(a-H)^2\right/(a-H-D)^2$ on the right-hand-side (again, $D$ is the ocean depth and $H$ is the ice thickness). By design, the system is in thermal balance and the global-mean ice shell thickness should not change with time.
  
  \subsection{Boundary conditions.}

  The heat generated in the silicate core is represented by an upward geothermal heat flux. According to \textit{Beuthe 2019}\cite{Beuthe-2019:enceladuss} and \textit{Choblet et al. 2017}\cite{Choblet-Tobie-Sotin-et-al-2017:powering}, the heat flux peaks at the two poles. We fit the meridional heat profile obtained by \textit{Beuthe 2019}\cite{Beuthe-2019:enceladuss} using a cosine function,
  \begin{equation}
    \label{eq:H-core}
    \mathcal{H}_{\mathrm{core}}(\phi)=\bar\mathcal{H}_{\mathrm{core}}\cdot(1.213-0.319\cos(2\phi)),
  \end{equation}
  where $\phi$ denotes latitude and $\bar\mathcal{H}_{\mathrm{core}}$ is the global mean heat flux from the bottom. The function inside the parenthesis is normalized to have a global average of unity. The profile of $\mathcal{H}_{\mathrm{core}}$ is plotted in Fig.~\ref{fig:schematics}b using a purple dashed curve (in the figure, $\bar\mathcal{H}_{\mathrm{core}}$ is set to be equal to $\mathcal{H}_{\mathrm{cond}}$, corresponding to the ``heat-from-below'' scenario). 
  
  The interaction between the ocean and ice is simulated using MITgcm's ``shelf-ice'' package \cite{Holland-Jenkins-1999:modeling}. We turn on the ``boundary layer'' option to avoid the numerical instability induced by an ocean layer which is too thin. The code is modified to account for a gravitational acceleration that is very different from that on earth, the temperature dependence of heat conductivity, and the meridional variation of tidal heating generated inside the ice shell and the ice surface temperature. As demonstrated in Fig.~\ref{fig:schematics}a, the freezing/melting rate of the ice shell is determined by a heat budget for a thin layer of ice at the base\footnote{This choice is supported by the fact that most tidal heating is generated close to the ocean-ice interface \cite{Beuthe-2018:enceladuss}.} (encompassed by dashed black lines), which involves three terms: the heat transmitted upward by ocean $\mathcal{H}_{\mathrm{ocn}}$, the heat loss through the ice shell due to heat conduction $\mathcal{H}_{\mathrm{cond}}$, and the tidal heating generated inside the ice shell $\mathcal{H}_{\mathrm{tidal}}$. Following \textit{Holland and Jenkins 1999}\cite{Holland-Jenkins-1999:modeling}, the continuity of heat flux and salt flux through the ``boundary layer'' gives,
  \begin{eqnarray}
    &~&\mathcal{H}_{\mathrm{ocn}}-\mathcal{H}_{\mathrm{cond}}+\mathcal{H}_{\mathrm{tidal}}=-L_fq\label{eq:boundary-heat}\\
&~&\rho_w\gamma_S(S_{\mathrm{ocn-top}}-S_b)=-qS_b, \label{eq:boundary-salinity}   
  \end{eqnarray}
  where $T_{\mathrm{ocn-top}}$ and $S_{\mathrm{ocn-top}}$ denote the temperature and salinity in the top grid of the ocean\footnote{When model resolution is smaller than the boundary layer thickness, the salinity below the upper-most grid cell also contributes to $T_{\mathrm{ocn-top}}$ and $S_{\mathrm{ocn-top}}$.}, $S_b$ denotes the and salinity in the ``boundary layer'', and $q$ denotes the freezing rate. $C_p$ is the heat capacity of the ocean, $L_f=334000$~J/kg is the latent heat of fusion of ice, and $\rho_w=1033$~kg/m$^3$ is the density of the ocean. $\gamma_S=5.05\times10^{-9}$~m/s is the exchange coefficients for salinity across the ``boundary layer''.
The expression of $\mathcal{H}_{\mathrm{cond}}$ is given in the previous section.  
The heat flux transmitted from the ocean into the ice shell $\mathcal{H}_{\mathrm{ocn}}$ can be written as
  \begin{equation}
    \label{eq:H-ocn}
    \mathcal{H}_{\mathrm{ocn}}=C_p\rho_w\gamma_T(T_{\mathrm{ocn-top}}-T_b),
  \end{equation}
  where $\gamma_T=10^{-6}$~m/s is the exchange coefficients for temperature and $T_b$ denotes the and temperature in the ``boundary layer''. $T_b$ equals the freezing temperature $T_m$ at pressure $P$ and salinity $S_b$ by definition. The following formula is used 
  \begin{equation}
    \label{eq:Tb}
    T_b=T_m(S_b,P)=c_0+b_0P+a_0S_b,
  \end{equation}
  where $P$ is the pressure under the ice shell, $a_0=-0.0575$~K/psu,~$b_0=-7.61\times10^{-4}$~K/dbar and $c_0=0.0901$~degC. 

  Assuming a globally-uniform surface temperature of 70K, we derive the ice properties, and thereby the tidal heating profile, following exactly the same procedure as in \textit{Kang \& Flierl 2020}\cite{Kang-Flierl-2020:spontaneous}, which is a simplified version of \textit{Beuthe 2019}\cite{Beuthe-2019:enceladuss}. The resultant $\mathcal{H}_{\mathrm{tidal}}$ profile is shown in Fig.~\ref{fig:schematics} by a red solid curve. Consistent with previous works \cite{Beuthe-2018:enceladuss}, the heating profile peaks at the two poles. The final expression is rather complicated, and thus is omitted for brevity. Interested readers are encouraged to read \cite{Kang-Flierl-2020:spontaneous} and \cite{Beuthe-2019:enceladuss}.

  The only two unknowns, $S_b$ and $q$, in Eq.~(\ref{eq:boundary-heat}) and Eq.~(\ref{eq:boundary-salinity}) can therefore be solved jointly. When freezing occurs ($q>0$), salinity flux $\rho_w\gamma_S(S_{\mathrm{ocn-top}}-S_b)$ is negative (downward). This leads to a positive tendency of salinity at the top of the model ocean. Broadly speaking\footnote{Some extra terms need to be included when the thickness of the top grid of the ocean is smaller than the boundary layer thickness.}, the tendency is 
  \begin{equation}
    \label{eq:S-tendency}
    \frac{dS_{\mathrm{ocn-top}}}{dt}=S_{b}\frac{q}{\delta z},
  \end{equation}
  where $\delta z$ is the thickness of the ``boundary layer'' at the ocean-ice interface.

The flow speed is relaxed to zero at the top and bottom boundaries with a rate of $2\times10^{-3}$~s$^{-1}$.
  
  \subsection{Initial conditions.}
  The total heat generation/loss is around 43~mW/m$^2$ (equivalent to 32GW of total dissipation rate). This is a tiny value, especially given the large heat capacity of the ocean. That means if the initial temperature is 10~mK warmer off the equilibrium, it would at least take around 1300 years to adjust through the heat surplus/shortage alone\footnote{This estimation does not consider the heat induced by phase change.}. In addition to the long equilibrium time, freezing and melting during the spin-up period may significantly change the salinity profile resulting in a circulation that would not otherwise be attainable. To obtain an initial temperature profile that is close to equilibrium, we first set  the temperature just under the ice shell ($T_{\mathrm{ocn-top}}$) so that $\mathcal{H}_{\mathrm{ocn}}=C_p\rho_w\gamma_T(T_{\mathrm{ocn-top}}-T_m(S_0,P))$ equals to the desired heat flux transmitted upward by the ocean. To transmit a heat flux of $\mathcal{H}_{\mathrm{ocn}}$ through the ocean, a vertical temperature gradient is required. We set the initial temperature profile to be a linear function of $z$, and adjust the slope to ensure the global mean salinity doesn't drift significantly in the initial $5$ model years. All simulations are run for 120~model years for ocean dynamics to approach equilibrium. 
    
\end{methods}

\bibliographystyle{naturemag}
\bibliography{export}

\begin{addendum}
 \item This work is carried out in the Department of Earth, Atmospheric and Planetary Science (EAPS) in MIT. WK acknowledges support as a Lorenz Fellow supported by endowed funds in EAPS. SB, JC, CS, CG, AT and JM acknowledge part-support from NASA Astrobiology Grant 80NSSC19K1427 “Exploring Ocean Worlds”.
 \item[Correspondence.] Correspondence and requests for materials should be addressed to Wanying Kang~(email: wanying@mit.edu).
\end{addendum}
\end{document}